\documentclass[prl,amsmath,amssymb,showpacs]{revtex4}
\usepackage{graphicx}

\begin{document}

\title{Regular and Black Hole Solutions in the Einstein-Skyrme Theory 
with Negative Cosmological Constant}

\author{Noriko Shiiki}
\email{norikoshiiki@mail.goo.ne.jp}
\author{Nobuyuki Sawado}
\email{sawado@ph.noda.tus.ac.jp}

\affiliation{Department of Physics, Tokyo University of Science, Noda, 
Chiba 278-8510, Japan}

\date{\today}


%%%%%%%%%%%%%%%%%%%%%%%%%%%%%%%%%%%%%%%%%%%
%             ABSTRACT                    %
%%%%%%%%%%%%%%%%%%%%%%%%%%%%%%%%%%%%%%%%%%%
\begin{abstract}
We study spherically symmetric regular and black hole solutions 
in the Einstein-Skyrme theory with a negative cosmological constant. 
The Skyrme field configuration depends on the value of the cosmological 
constant in a similar manner to effectively varying the gravitational 
constant. We find the maximum value of the cosmological constant above  
which there exists no solution. 
The properties of the solutions are discussed in comparison with the 
asymptotically flat solutions.  
The stability is investigated in detail by solving the linearly 
perturbed equation numerically. 
We show that there exists a critical value of the cosmological constant 
above which the solution in the branch representing unstable configuration 
in the asymptotically flat spacetime turns to be linearly stable. 
\end{abstract}

\maketitle 

\section{1. Introduction}
The Skyrme model is a unified theory of hadrons  
proposed by Skyrme~\cite{skyrme58}. 
The model consists of meson fields alone which are represented 
in terms of angular variables to be multi-valued. 
Associated with this non-linearity, a topological soliton solution 
called skyrmion arises and its topologically conserved charge is 
interpreted as the number of particle sources, baryon number $B$.  
Skyrme discussed the particle nature of a $B=1$ spherically symmetric 
skyrmion by imposing the hedgehog ansatz on the pion fields.  
Witten showed that QCD in the large-$N_{c}$ limit 
reduces to an effective theory of mesons, and baryons emerge as 
solitons in this weakly coupled meson theory~\cite{witten79}.  
The detailed analysis for the property of the $B=1$ skyrmion as 
a nucleon was performed in Ref.~\cite{adkins83} upon quantization 
of the collective coordinate. Axially symmetric $B=2$ skyrmions 
were found and quantised in Refs.~\cite{kopelio87,braaten88}. 
The ground state of the solution was shown to have the correct quantum 
numbers of the deuteron. Remarkably, multi-skyrmions with $B>2$ possess various
discrete symmetries analogously to multi-BPS monopoles~\cite{braaten90,houghton98}.

It has been known that the Einstein-Skyrme (ES) system possesses 
regular and black hole solutions. 
The spherically symmetric black hole solution with Skyrme 
hair~\cite{luckock86,luckock87,droz91,bizon92} and self-gravitating 
skyrmion~\cite{bizon92} were investigated. 
It was shown that there exist two fundamental branches 
of the solutions and interestingly one of the branches represents 
stable configuration under linear perturbations~\cite{heusler92,bizon92}. 
$B=2$ Skyrme black hole and regular solutions with axisymmetry were  
constructed in Ref.~\cite{shiiki04}. The review of the black hole 
solutions with Skyrme hair is given in Ref.~\cite{shiiki05}.  
All of these solutions are, however, constructed in the asymptotically flat  
spacetime. Recently in Ref.~\cite{shiiki05-1} we considered the Einstein-Skyrme 
system with a negative cosmological constant and found $B=1$ asymptotically 
anti-de Sitter (AdS) black hole solutions. 
In this paper we develop the previous work of Ref.~\cite{shiiki05-1} and study 
$B=1$ spherically symmetric regular and black hole solutions in the 
asymptotically AdS spacetime. 

In the context of the Einstein-Yang-Mills (EYM) theory, 
it was shown that regular and black hole solutions are unstable for 
$\Lambda \ge 0$~\cite{zhou90,torii95,volkov96}, but there exist stable 
solutions for $\Lambda < 0$~\cite{winstanley99,bjoraker00,breitenlohner04}. 
We investigate the linear stability of our solutions and discuss in detail 
to see if the presence of the cosmological constant changes the stability 
properties of the solutions as in the EYM theory.  

There has been an increasing interest in the AdS spacetime. 
Especially the AdS black hole is an interesting object from the 
holographic point of view in the form of AdS/CFT 
correspondence~\cite{maldacena98,witten98}.
Brane world cosmology also indicates that there was a period when  
spacetime was AdS with a negative cosmological constant in the 
early universe (for example, see~\cite{randall99,kaloper99,brevik00,karch01}). 
The solutions we obtain in this paper provide a semiclassical framework 
to study the interaction of a baryon and gravity or a primordial black hole 
with a negative cosmological constant. 

\section{2. The Einstein-Skyrme Model}
The Einstein-Skyrme system with a cosmological constant $\Lambda$ 
is defined by the action
\begin{eqnarray}
	{\cal S} = \int d^{4}x \sqrt{-g}\left\{ \frac{1}{16\pi G}(R-2\Lambda) 
	+\frac{F_{\pi}^{2}}{16}g^{\mu\nu}{\rm tr}
	\,(L_{\mu}L_{\nu})+\frac{1}{32e^{2}}g^{\mu\nu}g^{\rho\sigma}{\rm tr}\,
	([L_{\mu},L_{\rho}][L_{\nu},L_{\sigma}])\right\} \label{action} 
\end{eqnarray} 
where $L_{\mu}=U^{\dagger}\partial_{\mu}U$ and $U$ is an $SU(2)$ chiral field. 
$F_{\pi}$ is the pion decay constant and $e$ is a dimensionless 
parameter. 

We require that the spacetime recovers the AdS solution at infinity 
and thus parameterize the metric as   
\begin{eqnarray}
	ds^{2}=-e^{2\delta(r)}C(r)dt^{2}+C(r)^{-1}dr^{2}
	+r^{2}(d\theta^{2}+\sin^{2}\theta d\varphi^{2})   \label{}
\end{eqnarray}
where  
\begin{eqnarray}
	C(r)=1-\frac{2Gm(r)}{r}-\frac{\Lambda r^{2}}{3} \,. \label{}
\end{eqnarray}  
The topology of AdS spacetime is $S^{1}\times R^{3}$ and hence the 
timelike curves are closed. This can be, however, unwinded if we consider  
the covering spacetime with topology $R^{4}$. 

The $B=1$ skyrmion can be obtained by imposing the hedgehog ansatz on the 
chiral field
\begin{eqnarray}
	U({\vec r})=\cos f(r)+i{\vec n}\cdot{\vec \tau}\sin f(r) \,. \label{}
\end{eqnarray}
Introducing the dimensionless variables  
\begin{eqnarray}
	 x=eF_{\pi}r \; , \;\;\;\;\;\mu(x)=eF_{\pi}G m(r) \;, 
	\;\;\;\;\;{\tilde \Lambda}=\Lambda/e^{2}F_{\pi}^{2}  \label{}
\end{eqnarray} 
with 
\begin{eqnarray}
	C(x)=1-\frac{2\mu(x)}{x}-\frac{{\tilde \Lambda}x^{2}}{3} \,, \label{cx}
\end{eqnarray}
one obtains the skyrmion energy as 
\begin{eqnarray}
	E_{S}&=&4\pi \frac{F_{\pi}}{e} \int \left\{\frac{1}{8}\left(Cf'^{2}
	+\frac{2\sin^{2}f}{x^{2}}\right)+\frac{\sin^{2}f}{2x^{2}}
	\left(2Cf'^{2}+\frac{\sin^{2}f}{x^{2}}\right)\right\}e^{\delta} x^{2} dx \\
	&=& 4\pi \frac{F_{\pi}}{e} \int \left(\frac{1}{8}Cuf'^{2} 
	+\frac{1}{4x^{2}}v\right)e^{\delta} dx \label{energy}
\end{eqnarray}
where we have defined $u=x^{2}+8\sin^{2}f$ and $v=\sin^{2}f(x^{2}+2\sin^{2}f)$.  
The prime denotes the derivative with respect to $x$. 
The covariant topological current is defined by  
\begin{eqnarray}
	B^{\mu}=-\frac{\epsilon^{\mu\nu\rho\sigma}}{24\pi^{2}}\frac{1}{\sqrt{-g}}
	{\rm tr}\left(U^{-1}\partial_{\nu}UU^{-1}\partial_{\rho}UU^{-1}
	\partial_{\sigma}U\right) 
	\label{topological_current} \label{baryon_current}
\end{eqnarray}
whose zeroth component corresponds to the baryon number density 
\begin{eqnarray}
	B^{0}=-\frac{1}{2\pi^{2}}e^{-\delta}\frac{f'\sin^{2}f}{r^{2}}\, . \label{}
\end{eqnarray}
We impose the boundary condition on the profile function as 
\begin{eqnarray}
	f(x) \rightarrow 0 \;\;\; {\rm as} \;\;\; r  \rightarrow \infty \,,  
	\label{boundary}
\end{eqnarray}
which ensures the total energy (\ref{energy}) to be finite. 
Then the baryon number becomes    
\begin{eqnarray}
	B=\int \sqrt{-g}\,B^{0} \, d^{3}x = -\frac{2}{\pi}\int_{f_{0}}^{0}
	\sin^{2}f df = \frac{1}{2\pi}(2f_{0}-\sin 2f_{0}) \, . \label{baryon-number}
\end{eqnarray}
For regular solutions, $f_{0}=f(0)$ should take $\pi$ in order for the baryon 
number to be one. For black hole solutions, $f_{0}=f(x_{h})$ with event horizon 
$x_{h}$ takes the value less than $\pi$, which means that the solution possesses 
fractional baryonic charge. 
In this case $f(x_{h})$ is a shooting parameter 
determined numerically so as to satisfy the desired asymptotical behavior 
in Eq.~(\ref{boundary}).  

The Einstein equations with a cosmological constant $\Lambda$ takes the form
\begin{eqnarray}
	R_{\mu\nu}-\frac{1}{2}Rg_{\mu\nu}+\Lambda g_{\mu\nu}
	=8\pi GT_{\mu\nu}  \label{}
\end{eqnarray}
which reads     
\begin{eqnarray}
	G_{00}=8\pi GT_{00}-\Lambda g_{00} & \rightarrow &
	1-C-C'x=\frac{\alpha}{4}\left(Cuf'^{2}+\frac{2v}{x^{2}}\right) 
	+{\tilde \Lambda}x^{2} \\
	G_{11}=8\pi GT_{11}-\Lambda g_{11} & \rightarrow &
	-1+C+\frac{(e^{2\delta}C)'}{e^{2\delta}}x=\frac{\alpha}{4}
	\left(Cuf'^{2}-\frac{2v}{x^{2}}\right)-{\tilde \Lambda}x^{2}  \label{}
\end{eqnarray}
where we have defined the coupling constant $\alpha=4\pi GF_{\pi}^{2}$. 
Consequently following two equations are obtained for the gravitational fields 
\begin{eqnarray}
       \delta'= \frac{\alpha}{4x}uf'^{2} \label{stab_d} \;,\;\;\;\;\;
	 -(Cx)'+1 = \frac{\alpha}{4}\left(Cuf'^{2}+\frac{2v}{x^{2}}\right)
	 +{\tilde \Lambda}x^{2} \,. 
\end{eqnarray}  
Taking variation of the static energy (\ref{energy}) with respect to the 
profile $f(x)$, one can get the field equation for matter 
\begin{eqnarray}
	f''= \frac{1}{e^{\delta}Cu}\left[-(e^{\delta}Cu)'f'
	+\left(4Cf'^{2}+1+\frac{4\sin^{2}f}{x^{2}}\right)
	e^{\delta}\sin 2f \right] \, .
\end{eqnarray}
Thus the coupled field equations to be solved are given by 
\begin{eqnarray}
	 \delta'&=& \frac{\alpha}{4x}uf'^{2} \label{d-eq}\\ 
	 \mu'&=& \frac{\alpha}{8}\left(Cuf'^{2}+\frac{2v}{x^{2}}\right) 
	 \label{mu-eq} \\
	 f''&=& \frac{1}{e^{\delta}Cu}\left[-(e^{\delta}Cu)'f'
	 +\left(4Cf'^{2}+1+\frac{4\sin^{2}f}{x^{2}}\right)
	 e^{\delta}\sin 2f \right] \, .\label{f-eq}
\end{eqnarray}

%**************** Section 3 **************************
\section{3. Boundary Conditions}
The boundary conditions for regular solutions are determined 
by expanding the functions around the origin $x=0$ and comparing 
the coefficients in each order of $x$ in the field equations~(\ref{d-eq})-(\ref{f-eq}). 
As a result, we find 
\begin{eqnarray}
      f &=& \pi +f_{1}x+O(x^{3}) \\
	\delta &=& \delta_{0}+\frac{\alpha}{8}f_{1}^{2}(1+8f_{1}^{2})x^{2}
	+O(x^{3}) \\
      \mu &=& \frac{\alpha}{8}f_{1}^{2}(1+4f_{1}^{2})x^{3}+O(x^{4}) \,, \label{} 
\end{eqnarray}
where $f_{1}$ and $\delta_{0}$ are shooting parameters. $f_{1}$ is chosen 
so as to satisfy Eq.~(\ref{boundary}) and $\delta_{0}$ is chosen so as to 
recover the AdS spacetime asymptotically, that is, $\delta(x)\rightarrow 0$ 
as $x \rightarrow \infty$. 

Similarly, in order to determine the boundary conditions on the regular 
event horizon, let us expand the fields around the horizon $x=x_{h}$ 
\begin{eqnarray}
      f &=& f_{h}+f_{1}(x-x_{h})+O((x-x_{h})^{2}) \\
	\delta &=& \delta_{h}+\delta_{1}(x-x_{h})+O((x-x_{h})^{2}) \\
      \mu &=& \frac{x_{h}}{2}-\frac{{\tilde \Lambda}x_{h}^{3}}{6}
      +\mu_{1}(x-x_{h})+O((x-x_{h})^{2}) \,. \label{} 
\end{eqnarray}
Inserting them into the field equations~(\ref{d-eq})-(\ref{f-eq}) and 
comparing the coefficients in each order of $(x-x_{h})$, one obtains
\begin{eqnarray}
      f_{1}&=&\frac{x_{h}^{2}+4\sin^{2}f_{h}}{x_{h}(x_{h}^{2}+8\sin^{2}f_{h})
      (1-2\mu_{1}-{\tilde \Lambda}x_{h}^{2})}\sin 2f_{h} \\
	\delta_{1}&=&\frac{\alpha}{4x_{h}}
	(x_{h}^{2}+8\sin^{2}f_{h}) f_{1}^{2} \\
	\mu_{1}&=&\frac{\alpha}{4}\left(1+\frac{2\sin^{2}f_{h}}{x_{h}^{2}}
      \right)\sin^{2}f_{h} \,, \label{}
\end{eqnarray}
where $f_{h}$ and $\delta_{h}$ are shooting parameters with the desired 
asymptotic behavior $f(x)$, $\delta(x) \rightarrow 0$ as $x\rightarrow \infty$. 

%******************** Section 4 ******************************
\section{4. Numerical Results}
\begin{figure}
\includegraphics[height=6.5cm, width=8.5cm]{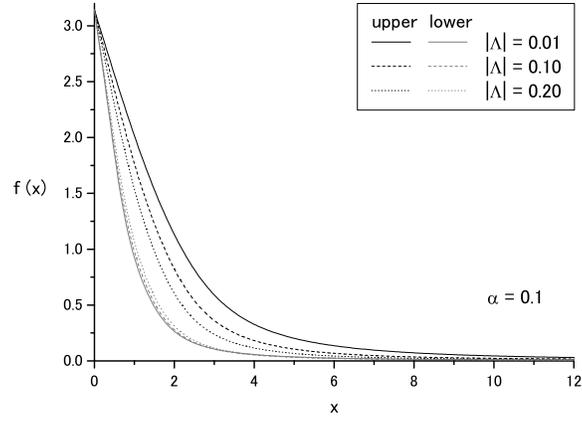}
\caption{\label{fig:f_reg_01} The profile function $f$ as a 
function of the radial coordinate $x$ for $|{\tilde \Lambda}| 
= 0.01, 0.1, 0.2$ with $\alpha=0.1$ fixed.}
\end{figure}

\begin{figure}
\includegraphics[height=6.5cm, width=8.5cm]{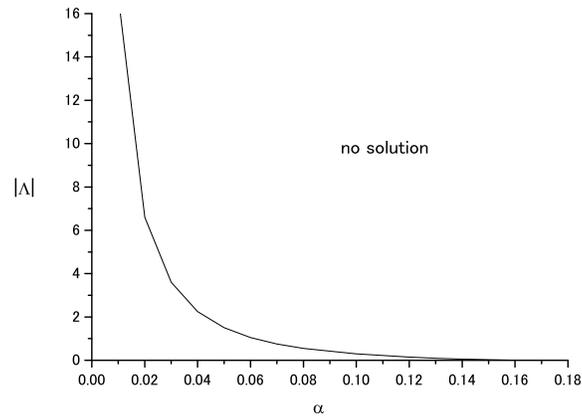}
\caption{\label{fig:a-L_reg} The solid line shows the $\alpha$ dependence 
of the maximum value of $|{\tilde \Lambda}|$ above which there exists no 
soliton solution.}
\end{figure}

\begin{figure}
\includegraphics[height=6.5cm, width=8.5cm]{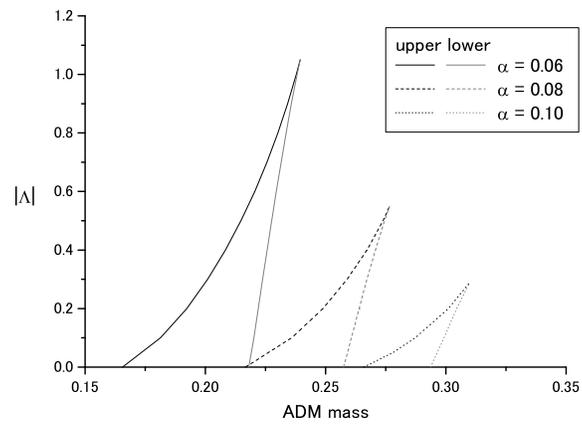}
\caption{\label{fig:Lambda-M_reg} The dependence of the ADM mass $M$ 
on the cosmological constant $|{\tilde \Lambda}|$ for $\alpha =0.06, 
0.08$ and $0.1$. }
\end{figure}

The profile functions of regular solutions are shown in Fig.~\ref{fig:f_reg_01} 
for several values of $|{\tilde \Lambda}|$ with $\alpha =0.1$ fixed. 
There are two branches of the solutions for each value of the cosmological 
constant as well as the coupling constant.  
We define the solution with larger value of $f_{1}$ as an upper branch 
and smaller value of $f_{1}$ as a lower branch. 
The skyrmion shrinks as $|{\tilde \Lambda}|$ becomes larger in the 
upper branch and expands slightly in the lower branch. 
Similar behavior is observed for the solutions in the asymptotically 
flat spacetime when $\alpha$ is increased~\cite{shiiki05}. 
Thus the variation of cosmological constant gives a similar effect 
on the skyrmion as the variation of the coupling constant does.  
We found the maximum value of the cosmological constant with $\alpha > 0$ 
above which there exists no solution. When $\alpha =0$, regular solutions 
exist for all values of ${\tilde \Lambda} \le 0$. The maximum value of 
$|{\tilde \Lambda}|$ is plotted as a function of $\alpha$ in 
Fig.~\ref{fig:a-L_reg}. $|{\tilde \Lambda}|$ decreases monotonically  
as $\alpha$ increases and at $\alpha \simeq 0.162$ only asymptotically 
flat spacetime solution exists. For $\alpha > 0.162$, we found no 
solution with $\Lambda \le 0$. Fig.~\ref{fig:Lambda-M_reg} shows 
the dependence of the ADM mass $M=\mu (\infty)$ on $\alpha$ and 
$|{\tilde \Lambda}|$. The ADM mass gives the total energy available 
in the spacetime and therefore, for regular solutions, it is equivalent to 
the skyrmion energy ${\tilde E}_{S}=\mu (\infty)$ where 
${\tilde E}_{S}=eE_{S}/F_{\pi}$ in Eq.~(\ref{energy}). 
As to be expected, the mass increases 
as $\alpha$ and/or $|{\tilde \Lambda}|$ increase. It is observed that 
the presence of the cosmological constant affects the upper-branch 
significantly more than the lower-branch. 
 
\begin{figure}
\includegraphics[height=6.5cm, width=8.5cm]{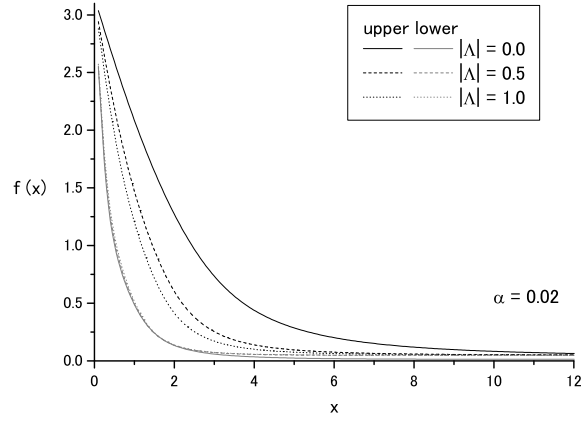}
\caption{\label{fig:F-a002} The profile function $f$ as a 
function of the radial coordinate $x$ for $|{\tilde \Lambda}| 
= 0.0, 0.5, 1.0$ with $x_{h}=0.1$ and $\alpha=0.02$ fixed.}
\end{figure}

For black hole solutions, the profile function numerically computed for 
several values of $|{\tilde \Lambda}|$ are shown in Fig.~\ref{fig:F-a002}. 
The horizon radius and the coupling constant are fixed with $x_{h}=0.1$ and 
$\alpha=0.02$. There are two branches of solutions for each value of the 
cosmological constant as was seen in the regular case. 
The dependence of profiles on the cosmological constant is also similar 
with the regular case. In the lower branch, however, the change in size 
is much smaller and is almost unrecognizable. 

\begin{figure}
\includegraphics[height=6.5cm, width=8.5cm]{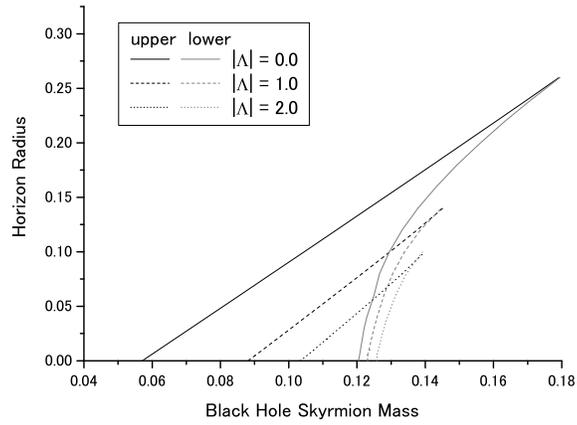}
\caption{\label{fig:Mbh-xh} The horizon radius $x_{h}$ 
as a function of the black hole skyrmion mass $M_{bh}$ for 
$|{\tilde \Lambda}|= 0.0, 1.0, 2.0$ with $\alpha=0.02$ fixed.}
\end{figure}

The black hole skyrmion mass-horizon radius relation is shown in 
Fig.~\ref{fig:Mbh-xh}. The skyrmion energy ${\tilde E}_{S}$ is related to 
the black hole skyrmion mass $M_{bh}=\mu(\infty)$ by 
\begin{eqnarray}
	{\tilde E}_{S}=M_{bh}-\mu (x_{h}) =\mu(\infty)-\frac{x_{h}}{2}\,. \label{}
\end{eqnarray}
Since the entropy of the black hole is written by 
\begin{eqnarray}
	S=\frac{\pi r_{h}^{2}}{4\hslash G}=\frac{\pi^{2}}{\hslash e^{2}}
	\left(\frac{x_{h}^{2}}{\alpha}\right)\,, \label{entropy}
\end{eqnarray} 
one can see that the upper and lower branch correspond to the high- and 
low-entropy branch respectively. The cosmological constant reduces the 
entropy of the black hole. The reduction of the entropy is also seen 
when the coupling constant increases as is inferred from Eq.~(\ref{entropy}). 

\begin{figure}
\includegraphics[height=6.5cm, width=8.5cm]{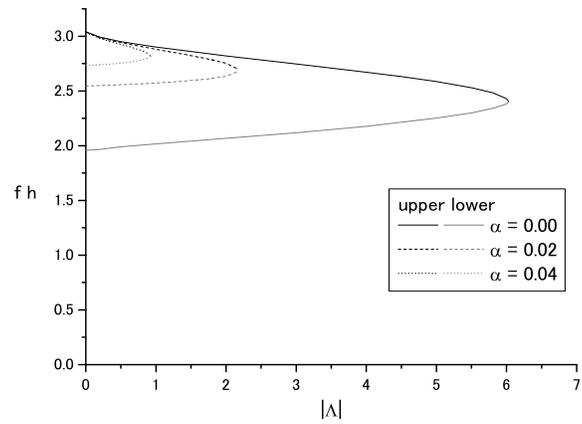}
\caption{\label{fig:L-fh} Shooting parameter $f_{h}$ as a function of 
$|{\tilde \Lambda}|$ for $\alpha = 0.0, 0.02, 0.04$.}
\end{figure}

Fig.~\ref{fig:L-fh} shows the parameter $f_{h}$ as a function of 
$|{\tilde \Lambda}|$ for $\alpha=0.0,\,0.02,\,0.04$ with $x_{h}=0.1$ fixed. 
The value of $f_{h}$ is directly related 
to the baryon number as can be seen from Eq.~(\ref{baryon-number}). 
Thus, in the upper branch, the baryon number becomes smaller  
as $|{\tilde \Lambda}|$ becomes larger, which represents the baryon more absorbed 
by the black hole. On the other hand, in the lower branch, the baryon number 
slightly increases as $|{\tilde \Lambda}|$ becomes larger. 
This result also shows that the cosmological constant gives a similar effect 
on the skyrmion as the coupling constant.

\begin{figure}
\includegraphics[height=6.2cm, width=8.5cm]{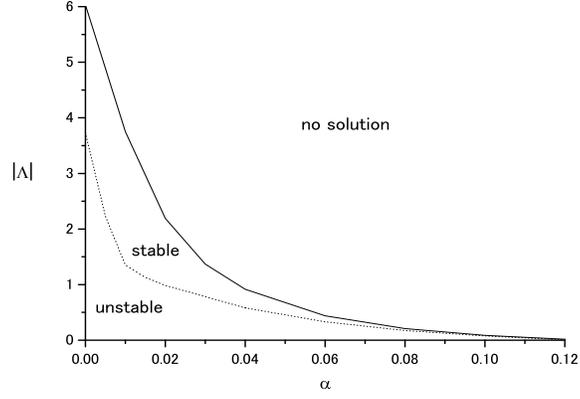}
\caption{\label{fig:a-L} The solid line shows the $\alpha$ dependence of the maximum 
value of $|{\tilde \Lambda}|$ above which there exists no black hole solution. 
The dotted line shows the $\alpha$ dependence of the value of $|{\tilde \Lambda}|$  
above which the lower branch solution change its stability to become stable.}
\end{figure}

We found the maximum value of $|{\tilde \Lambda}|$ above which there 
exists no black hole solution for each value of the coupling constant.  
In Fig.~\ref{fig:a-L}, the maximum value of $|{\tilde \Lambda}|$ is shown 
as a function of $\alpha$. The maximum value decreases monotonically as $\alpha$ 
increases, and at $\alpha=0.126$ it becomes zero. Thus at $\alpha=0.126$, 
the asymptotically AdS solution does not exist and 
only the asymptotically flat solution exists. For $\alpha >0.126$, we found 
no solution with ${\tilde \Lambda}\le 0$.     

Let us denote that for both the regular and black hole cases, the lower and upper 
branch solution coalesces at the maximum value of $|{\tilde \Lambda}|$. 
 
\section{5. Linear Stability Analysis}

In this section we shall examine the linear stability of the solutions 
described in the previous section. 
Let us consider the time-dependent small fluctuation around the 
static classical solutions $f_{0}$, $\delta_{0}$ and $\mu_{0}$ 
\begin{eqnarray}
	f(r,t) &=& f_{0}(r) +f_{1}(r,t) \label{f1} \\
	\delta (r,t)&=& \delta_{0}(r)+\delta_{1}(r,t) \label{d1} \\
	\mu (r,t)&=& \mu_{0}(r)+\mu_{1}(r,t)\,. \label{m1}
\end{eqnarray}
From the time-dependent Einstein-Skyrme action  
\begin{eqnarray}
	{\cal S}=-\frac{\pi e^{2}F_{\pi}^{4}}{2}\int \left[(-\frac{1}{e^{\delta}C}
	{\dot f}^{2}+Cf'^{2})u + v \right]e^{\delta} dx \, , \label{time-action}
\end{eqnarray}
one obtains the time-dependent field equation as  
\begin{eqnarray}
	(e^{\delta}Cuf')'+\frac{1}{2}\left(\frac{1}{e^{\delta}C}{\dot f}^{2}
	-e^{\delta}Cf'^{2}\right)u_{f}-\frac{e^{\delta}v_{f}}{x^{2}}
	=\frac{1}{e^{\delta}C}u{\ddot f} \label{time-field-eq}
\end{eqnarray}
where we have defined $u_{f}=\delta u/\delta f$ and $v_{f}=\delta v/\delta f$. 
The dot denotes the time derivative.  

The time-dependent Einstein equations are then given by  
\begin{eqnarray}
	G_{00}=8\pi GT_{00} & \rightarrow &
	1-C-C'x=\frac{\alpha}{4}\left[\left(\frac{1}{e^{2\delta}C}{\dot f}^{2}
	+Cf'^{2}\right)u+\frac{2v}{x^{2}}\right]\\
	G_{11}=8\pi GT_{11} & \rightarrow &
	-1+C+\frac{(e^{2\delta}C)'}{e^{2\delta}}x=\frac{\alpha}{4}
	\left[\left(\frac{1}{e^{2\delta}C}{\dot f}^{2}+Cf'^{2}\right)u
	-\frac{2v}{x^{2}}\right] \label{}
\end{eqnarray}
which reads the following two equations  
\begin{eqnarray}
	&& \delta'= \frac{\alpha}{4x}\left(\frac{1}{e^{2\delta}C^{2}}{\dot f}^{2}
	+f'^{2}\right)u \label{stab_d} \\ 
	&& -(Cx)'+1 = \frac{\alpha}{2x^{2}}v+C\delta'x \label{stab_c} \,.\label{}
\end{eqnarray}
Substituting Eqs.~(\ref{f1})-(\ref{m1}) into Eqs.~(\ref{stab_d}) and (\ref{stab_c}) 
gives the linearized equations   
\begin{eqnarray}
        \delta_{1}' &=& \frac{\alpha}{2x}(2u_{0}f_{0}'f_{1}'
       +u_{f_{0}}f_{0}'^{2}f_{1}) \label{ddel1} \\
	  -(e^{\delta_{0}}C_{1}x)'&=& \frac{\alpha}{2x^{2}}e^{\delta_{0}}
	v_{f_{0}}f_{1}+e^{\delta_{0}}C_{0}\delta_{1}'x \, .\label{dc1}
\end{eqnarray}
Eq.~(\ref{ddel1}) and the classical field equation Eq.~(\ref{time-field-eq}) 
which can be rewritten as  
\begin{eqnarray}
	 \frac{e^{\delta_{0}}v_{f_{0}}}{x^{2}}=(e^{\delta_{0}}C_{0}u_{0}f_{0}')'
	-\frac{1}{2}e^{\delta_{0}}C_{0}u_{f_{0}}f_{0}'^{2}\, , \label{static-field}
\end{eqnarray}
are inserted into Eq.~(\ref{dc1}) and resultantly one gets  
\begin{eqnarray}
	-(e^{\delta_{0}}C_{1}x)'=\frac{\alpha}{2}(e^{\delta_{0}}
	C_{0}u_{0}f_{0}'f_{1})' \,.  \label{}
\end{eqnarray}
This equation can be integrated immediately to obtain 
\begin{eqnarray}
	C_{1}=-\frac{\alpha}{2x}C_{0}u_{0}f_{0}'f_{1} \,. \label{c1}
\end{eqnarray}
Similarly let us linearize the field equation~(\ref{time-field-eq}). 
Using Eqs.~(\ref{ddel1}), (\ref{static-field}) and (\ref{c1}), one arrives at   
\begin{eqnarray}
	(e^{\delta_{0}}C_{0}u_{0}f_{1}')'-U_{0}f_{1}=\frac{1}{e^{\delta_{0}}C_{0}}
	u_{0}{\ddot f_{1}} \label{f1-eq}
\end{eqnarray}
where 
\begin{eqnarray}
	U_{0}&=&-(e^{\delta_{0}}C_{0}u_{f_{0}}f_{0}')'
	+\left(\frac{\alpha}{2x}e^{\delta_{0}}C_{0}u_{0}^{2}
	f_{0}'^{2}\right)'-\frac{\alpha}{2x}e^{\delta_{0}}C_{0}
	u_{0}u_{f_{0}}f_{0}'^{3} \nonumber \\
	&& +\frac{1}{2}e^{\delta_{0}}C_{0}u_{ff_{0}}f_{0}'^{2}
	+\frac{e^{\delta_{0}}v_{ff_{0}}}{x^{2}}\,. \label{}
\end{eqnarray}
Setting $f_{1}=\xi (x)e^{i\omega t}/\sqrt{u_{0}}$ , we derive from Eq.~(\ref{f1-eq}) as  
\begin{eqnarray}
	-(e^{\delta_{0}}C_{0}\xi')'+\left[\frac{1}{2\sqrt{u_{0}}}
	\left(e^{\delta_{0}}C_{0}\frac{u_{0}'}{\sqrt{u_{0}}}\right)'
	+\frac{1}{u_{0}}U_{0}\right] \xi =\omega^{2}
	\frac{1}{e^{\delta_{0}}C_{0}}\xi \,. \label{eigen-eq}
\end{eqnarray}
Let us introduce the tortoise coordinate $x^{*}$ such that 
\begin{eqnarray}
	\frac{dx^{*}}{dx}=\frac{1}{e^{\delta_{0}}C_{0}} \label{}
\end{eqnarray}
with $-\infty < x^{*} < +\infty$. Eq.~(\ref{eigen-eq}) is then reduced 
to the Strum-Liouville equation 
\begin{eqnarray}
	-\frac{d^{2}\xi}{dx^{*2}}+{\hat U}_{0}\xi =\omega^{2}\xi  \label{}
\end{eqnarray}
where
\begin{eqnarray}
	{\hat U}_{0}=e^{\delta_{0}}C_{0}\left[\frac{1}{2\sqrt{u_{0}}}
	\left(e^{\delta_{0}}C_{0}\frac{u_{0}'}{\sqrt{u_{0}}}\right)'
	+\frac{1}{u_{0}}U_{0} \right]\,. \label{strum}
\end{eqnarray}
The classical solution is linearly stable if there exists no negative eigenvalue  
since imaginary $\omega$ represent exponentially growing modes. 
Unfortunately the potential ${\hat U}_{0}$ has a complicated 
form and we are unable to discuss the stability analytically.  
Thus we solve the wave equation (\ref{eigen-eq}) numerically under 
the boundary conditions that $\xi$ vanishes at the boundaries, 
which ensure the norm of the wave function to be finite. 
The ground state corresponds to the wave function with no node. 
The $n$th excited state corresponds to the wave function with $n$ nodes. 

\begin{figure}
\includegraphics[height=6.2cm, width=8.5cm]{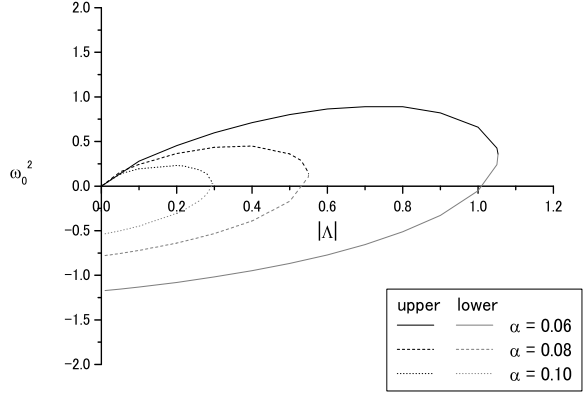}
\caption{\label{fig:L-omega_reg} Regular : The dependence of the ground state 
energy $\omega_{0}^{2}$ on the cosmological constant $|{\tilde \Lambda}|$ and 
the coupling constant $\alpha$. }
\end{figure}

\begin{figure}
\includegraphics[height=6.2cm, width=8.5cm]{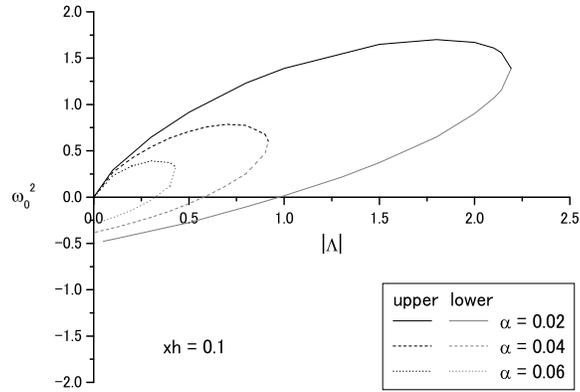}
\caption{\label{fig:L-omega_bh} Black Hole : The dependence of the ground state 
energy $\omega_{0}^{2}$ on the cosmological constant $|{\tilde \Lambda}|$ 
and the coupling constant $\alpha$ with $x_{h}=0.1$ fixed. }
\end{figure}

We show the eigenvalue of the ground state as a function of $|{\tilde \Lambda}|$ 
in Fig.~\ref{fig:L-omega_reg} for regular solutions and in 
Fig.~\ref{fig:L-omega_bh} for black hole solutions. 
For all values of $|{\tilde \Lambda}|$, upper branch solutions have  
positive eigenvalues, and hence they are stable. 
Remarkably the eigenvalues of lower branch solutions increase as 
$|{\tilde \Lambda}|$ increases, and at some value, they cross zero 
to become positive. These figures show clearly how the cosmological constant 
tilts the eigenvalues to the positive direction. 
Thus lower branch solutions change their stability at the critical 
value of the cosmological constant.  
The stability of black hole solutions more strongly depends 
on the cosmological constant than that of regular solutions.  

The results we have obtained indicate that the presence of a negative 
cosmological constant stabilizes both of regular and black hole solutions. 
This is consistent with the study of the EYM system where it was 
shown that the EYM solution is unstable for 
$\Lambda \ge 0$~\cite{zhou90,torii95,volkov96}, but 
the stable EYM solution exists for $\Lambda <0$~\cite{winstanley99,
breitenlohner04}. 

Another important feature for the ES solutions with $\Lambda <0$ 
is that only discrete modes exist in both of the branches. 
It is because the potential behaves asymptotically as $x^{2}$, meaning 
all the eigenvalues are discretized analogous to the harmonic oscillator 
eigenvalues. 
 
\section{6. Conclusions}

We have studied regular and black hole solutions in the Einstein-Skyrme system 
with a negative cosmological constant. There exist two fundamental branches 
of the solutions. For black holes, these corresponds to the high- and low-entropy 
branch respectively. 
The increase in the absolute value of the cosmological constant 
$|\Lambda|$ gives similar effects on the skyrmion as increasing effectively 
the value of the gravitational constant.  
The skyrmion shrinks in the upper-branch and expands in the lower-branch 
as $|\Lambda|$ increases. Particularly, in the black hole case the baryon 
number is more absorbed by the black hole corresponding to the increase 
in $|\Lambda|$.  
There is the maximum value of $|\Lambda|$ above which no solution 
exists for each value of the coupling constant. 
We have observed that the critical value of $\alpha$ for black hole 
solutions to exist is $0.126$ and $0.162$ for the regular case. 
 
The linear stability was examined in detail by solving the linear perturbed 
wave equation numerically. 
In the asymptotically flat case, it was shown that the upper branch is stable 
and the lower branch is unstable~\cite{heusler92,bizon92}. 
In the AdS case, however, there exist stable solutions 
even in the lower branch depending on the value of $|\Lambda|$. 
We have shown by numerically solving the linearly perturbed equation 
that the presence of a negative cosmological constant lifts 
the eigenvalues from negative to positive values. 
The observation that the negative cosmological constant stabilizes the ES  
solutions is consistent with the results in the EYM theory where 
stable regular and black hole solutions exist only when 
$\Lambda <0$~\cite{winstanley99,bjoraker00,breitenlohner04}.   
Let us give a brief comment on the catastrophe theory applied 
to the stability analysis of non-abelian black holes in 
Ref.~\cite{maeda94}. It seems that for asymptotically non-flat spacetimes, 
the catastrophe theory is not applicable to the stability analysis of 
solutions. Although concrete analysis should be performed for any statement 
on this matter to be confirmed, we suspect that the cosmological constant 
needs to be taken into account as an additional control parameter in the 
parameter space, which makes the dimension of the Whitney surface three with  
$S=S(M,B_{H},\Lambda)$ in the notation of Ref.~\cite{maeda94}. 
This extended catastrophe theory may be applicable for spacetimes with $\Lambda$. 
 
The solutions exhibited in this paper provide a semiclassical 
framework to study the interaction of a baryon and gravity or a primordial 
black hole in the presence of a negative cosmological constant. 
If the universe had gone through the AdS phase in the early epoch 
as indicated in Refs.~\cite{randall99,kaloper99,brevik00,karch01}, 
those solutions may have been produced after the hadronization. 
Since our model predicts the baryon decay by primordial black holes, 
the Einstein-Skyrme theory should be useful as a simple framework to study 
such decay process. 
 
The Skyrme model corresponds to QCD in the large $N$ limit and therefore  
it may be worth understanding the Skyrme model in the 
context of string and brane theories. 
Also interesting applications are to find solutions with nonspherical event 
horizon~\cite{bij02} or with $B=2$ axial symmetry~\cite{radu04} which 
were already discovered in the EYM theory. 

Finally, the numerical method employed to integrate the differential 
equations is based on the fourth-order Runge-Kutta method with a  
grid size $0.05$.

\pagebreak 

\end{document}